 \documentclass[preprint2]{aastex}

\shorttitle{Porosities of Dust Agglomerates}
\shortauthors{Teiser et al.}

\begin{document}


\title{Porosities of Protoplanetary Dust Agglomerates from Collision Experiments}


\author{J. Teiser\altaffilmark{1}, I. Engelhardt\altaffilmark{1} and G. Wurm \altaffilmark{1}}
\affil{Faculty of Physics, University Duisburg-Essen,
    Duisburg, Germany}

\email{jens.teiser@uni-due.de}






\begin{abstract}
Aggregation of dust through sticking collisions is the first step of planet formation. Basic physical properties of the evolving dust aggregates strongly depend on the porosity of the aggregates, e.g.  mechanical strength, thermal conductivity, gas-grain coupling time. Also the outcome of further collisions depends on the porosity of the colliding aggregates. In laboratory experiments we study the growth of large aggregates of $\sim$ 3 mm to 3 cm through continuous impacts of small dust agglomerates of 100 $\mu$m size, consisting of $\mu$m grains at different impact velocities. The experiments show that agglomerates grow by direct sticking as well as gravitational reaccretion. The latter can be regarded as suitable analog to reaccretion of fragments by gas drag in protoplanetary disks. Experiments were carried out in the velocity range between 1.5 m/s and 7 m/s. With increasing impact velocities the volume filling factor of the resulting agglomerates increases from $\phi = 0.2$ for 1.5 m/s to $\phi = 0.32$ for 7 m/s. These values are independent of the target size. Extrapolation of the measured velocity dependence of the volume filling factor implies that higher collision velocities will not lead to more compact aggregates. Therefore, $\phi = 0.32$ marks a degree of compaction suitable to describe structures forming at $\rm v > 6\, m/s$. At small collision velocities below 1 m/s highly porous structures with $\phi \approx 0.10$ will form. For intermediate collision velocities porosities vary. Depending on the disk model and resulting relative
velocities, objects in protoplanetary disks up to dm-size might evolve from highly porous ($\phi \approx 0.10$)  to compact ($\phi = 0.32$) with a more complex intermediate size range of varying porosity.

\end{abstract}



\section{Introduction}

Coagulation of dust particles by mutual collisions is a fundamental process occuring in protoplanetary disks. It marks the  first steps of planet formation. The formation of millimeter size aggregates from micron-sized dust particles proceeds rapidly. As collision velocities in this range are small and particles stick together easily, the growth of aggregates is indisputable \citep{Blum2008, Dominik1997}. With increasing particle size, also the relative velocities between particles increase and velocities reach values up to 60 m/s for meter size bodies or even more, eventually \citep{Weidenschilling1993, Brauer2008}. With increasing velocities collision results become more complex and compaction, fragmentation and rebound are typical outcomes \citep{Guttler2010, Zsom2010}. How the formation of larger objects proceeds from here is not settled yet. One way might be a continuous process of sticking collisions. Experimental studies show that under certain conditions growth is still possible even in the high velocity range with velocities of up to 60 m/s \citep{Teiser2009a, Wurm2005b}. In fact fragmentation assists growth at high speed impacts, as a large amount of the impact energy is dissipated by fragmentation. The experiments were -- somewhat arbitrarily -- based on collisions into compact targets with volume filling factors of 33\%.
The volume filling factor is defined as the ratio between the volume filled by dust to the total volume of a dust aggregate or $\phi = \rho_{agg} / \rho_{material}$ where $\rho_{agg}$ is the density of the aggregate and $\rho_{material}$ is the density of the used material. However, the mechanical properties of the target bodies determine the outcome of a collision and the mechanical properties of dust agglomerates are tied to the volume filling factor of the aggregates \citep{Beitz2011, Wurm2005a, Paraskov2007}. 

Beyond direct sticking, several authors propose different instability and gravity driven trapping and concentration scenarios to form planetesimals or even larger bodies \citep{Johansen2007, Johansen2007a, Youdin2007}.  These mechanisms strongly depend on the gas-grain coupling times of large aggregates. The gas-grain coupling time depends on the mass over surface ratio. That means that  the particle size for which these mechanisms work depends on the porosity, so porosity is also important in this context. 

The basic question behind this work therefore is how porosities evolve through the collisional history of growing aggregates. 
Initially dust growth proceeds as fractal growth \citep{Blum2008, Blum2000, Wada2008, Suyama2008}. With increasing particle size the collision velocities increase and particles become compacted during collisions \citep{Blum2000, Weidling2009, Paraskov2007, Kothe2010}. According to \citet{Weidenschilling1997}, decimeter bodies gain most of their mass by accretion of small (50 to 100 $\mathrm{\mu}$m) particles at collision velocities of about \mbox{10 m/s}. Therefore, the basic process
which we simulate in laboratory experiments, reported here, is that a body grows by continuous impacts of aggregates in this size range $(\rm \sim 100\mu m)$.

In a first experimental study of this kind \citet{Teiser2009b} investigated the structure of evolving decimeter size bodies, which form by multiple impacts of small particles at collision velocities of 7.7 m/s. They showed that dust agglomerates forming by direct sticking and reaccretion have a constant volume filling factor of $\phi = 0.31 \pm 0.02$. The collision history is erased during this growth. Independent of the special morphology of the surface after an impact, the top layer is compacted to the same level by following impacts for targets of mm to cm size. 
The experiments by \citet{Teiser2009b} were performed with one collision velocity only to determine the properties of decimeter aggregates. Smaller bodies are expected to form at lower impact speed though and the porosities might evolve differently. Just how the dependency of the volume filling factor on collision velocity looks like has been unclear. 
\citet{Kothe2010} found a rather strong increase of the volume filling factor with collision velocity for multiple impacts. 
However, their projectiles were larger (mm), highly porous, did always hit the same spot and consisted of spherical particles. 
This situation is different for aggregates building a surface in a random way through collisions as in \citet{Teiser2009b}. 
Within this work we present experiments in analogy to our earlier study in \citet{Teiser2009b}  forming bodies by multiple impacts of small particles but for varying velocities, which we regard as a possible scenario how macroscopic bodies evolve by accretion of small particles at typical collision speeds between 1.5 m/s and \mbox{7 m/s}.  

\section{Experimental aspects}

\subsection{Methods}\label{methods}

\begin{figure}[tb]
\center\includegraphics[width=5.5cm]{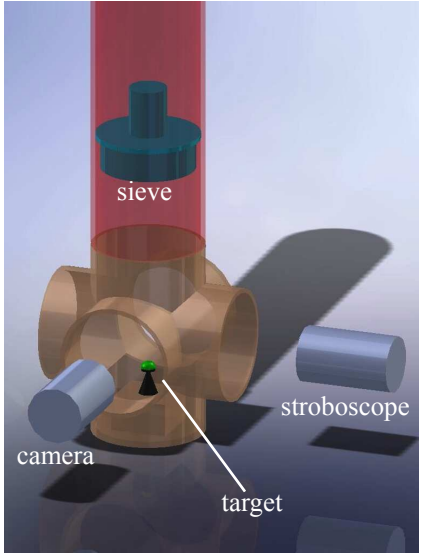}
\caption{Experimental setup for multiple collisions at impact velocities between \mbox{1.5 m/s} and \mbox{7 m/s.}}
\label{setup}
\end{figure}

The experiments are carried out with a similar method as the experiments already described by \citet{Teiser2009b}. Small compact dust agglomerates ($\sim 33 \%$ volume filling factor) collide with solid targets. The experimental setup is shown in Fig. \ref{setup}. The central part of the experiment is a sieve with a mesh width of \mbox{250 $\mu$m} and a rotating stirrer inside. Due to the stirrer a constant beam of dust agglomerates leaves the sieve and is accelerated by gravity. The experiments are carried out at pressures below \mbox{$10^{-2}$ mbar} to neglect the influence of gas drag. The drop height is adjustable between \mbox{0.1 m} and \mbox{2.5 m}, which leads to impact velocities between \mbox{1.5 m/s} and \mbox{7 m/s}. 

The collisions are observed with a video camera (25 frames/s) and illuminated by a stroboscope at a flash frequency of \mbox{500 Hz.}  Each frame therefore is a multiple exposure, so the particle trajectories are directly visible. From the distance between two particle images the impact velocities can be derived directly with the known time interval between two flashes of the stroboscope. Due to the launch mechanism the projectiles have vertical trajectories. The impact ejecta leave the target under varying angles. The horizontal $(v_x)$ and the vertical $(v_z)$ component of the ejecta velocities are determined by fitting parabolas to the trajectories. Due to the vacuum only gravity ($g$) acts on the ejected particles. 

After the first impacts a dust cover forms on any arbitrary substrate used as target. All following impacts are pure dust on dust collisions. Targets grow by a combination of direct sticking and reaccretion of ejecta due to gravity. The latter can be regarded as analog for gas driven reaccretion in protoplanetary disks \citep{Wurm2001a, Wurm2001b}. There, collisions between large dust aggregates (few cm or larger) with small ($\sim 100 \mu$m) particles are induced by their different coupling to the surrounding gas. Impact ejecta therefore feel the relative motion of the larger body with respect to the gas as head wind, which drives them back to the target surface. \citet{Teiser2009b} showed that gravity in the laboratory simulates the acceleration of impact fragments by the gas drag very well, as it is of the same order. While details of accretion efficiencies depend on the target size and gas density the effect will also be present in protoplanetary disks.
In the experiments we used metal cylinders of eight different sizes as targets. Therefore, ejecta of a given velocity might be reaccreted on the large targets but not on the smallest one. This way, we get a variation in the ratio between direct sticking and reaccretion of ejecta. The target sizes range from 3.2 mm to 36.6 mm in diameter. 

\begin{figure}[tb]
\center\includegraphics[width=7.0cm]{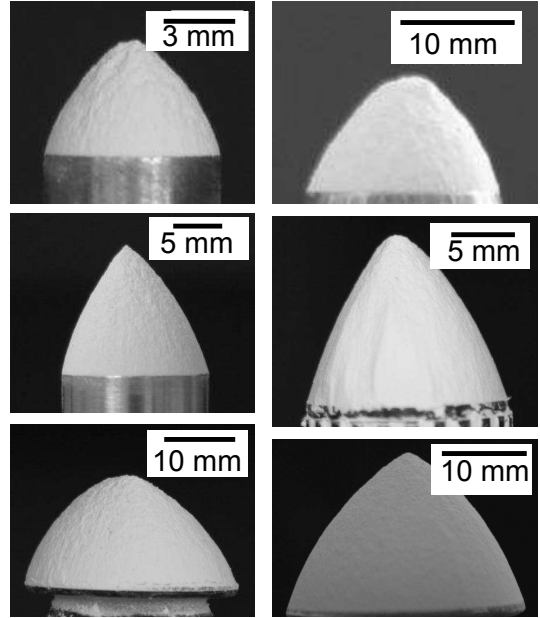}
\caption{Examples for targets of different sizes, which formed at 5.3 m/s (left column) and at 6.9 m/s (right column).}
\label{targets}
\end{figure}

The growing targets are very similar in their shape. A collection of targets of different sizes and grown at different impact velocities is shown in Fig. \ref{targets}. The targets can be described as cones with their slopes decreasing to the top. To determine the volume from the images, the projected surface is analyzed.  It is assumed that each target consists of circular disks stacked upon each other, which might be shifted with respect to each other. With this assumption and with $n$ circular disks with a diameter $d_n$ and a thickness $D$, the volume of the target can be calculated via

\begin{equation}
V = \pi D \cdot \sum_n \frac{d_n^2}{4}\,.
\end{equation}

For the analysis of the images, the thickness of the disks is set to one pixel. For each target several images from different sides are analyzed and this method is repeated for each image and for all eight different targets. 

\subsection{Parameters}\label{parameters}

All experiments described in this work are carried out with irregularly shaped quartz dust. The sizes of single monomers range from \mbox{0.1 $\mu$m} to \mbox{10 $\mu$m} with \mbox{$80 \%$} of the mass being between \mbox{1 $\mu$m} and \mbox{5 $\mu$m}. The same dust material has already been used in numerous experimental studies on protoplanetary growth \citep{Teiser2009a, Teiser2009b, Wurm2005b}.

\begin{figure}[tb]
\includegraphics[width=7.5cm]{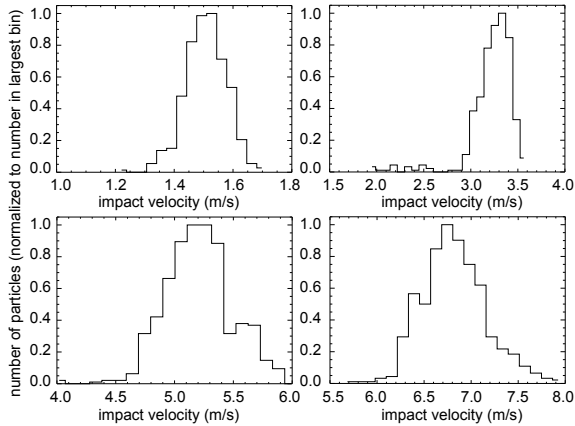}
\caption{Velocitiy distributions of particles at different drop heights.}
\label{velocities}
\end{figure}

The particles are launched with a sieve with a mesh width of \mbox{250 $\mu$m}, which is the same mesh width used by \citet{Teiser2009b}. Due to the dimensions of the drop tube, the drop height can be adjusted between 0.1 m and 2.5 m, which leads to impact velocities between 1.5 m/s and 7 m/s. The projectiles have varying initial velocities due to the launching process. Within this study four different drop heights (0.1 m, 0.5 m, 1.5 m, and 2.5 m) and therefore four impact velocities have been analyzed. Fig. \ref{velocities} shows the velocity distributions measured for the different drop heights. The mean velocities and their standard deviations have been determined as \mbox{$1.511$ m/s}  \mbox{$\pm\, 0.075$ m/s}, \mbox{$3.222$ m/s}  \mbox{$\pm\, 0.245$ m/s}, \mbox{$5.327$ m/s} \mbox{$\pm\, 0.299$ m/s}, and \mbox{$6.939$ m/s}  \mbox{$\pm\, 0.361$ m/s} for increasing drop heights, respectively. 

\section{Results}\label{results}

\subsection{Coefficient of restitution} \label{rest}

\begin{figure}[tb]
\includegraphics[width=7.5cm]{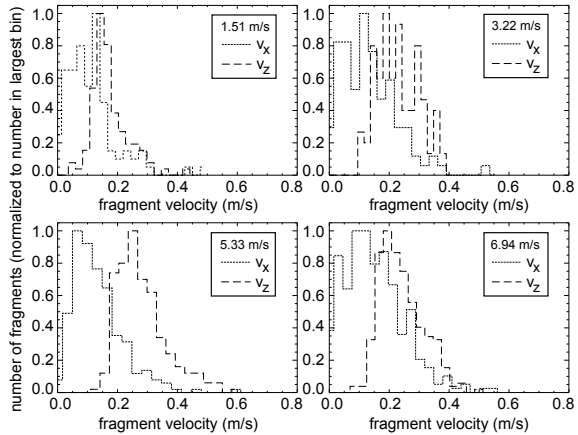}
\caption{Velocitiy distributions of impact ejecta for the studied collision velocities.}
\label{ejecta}
\end{figure}

All collisions lead to a combination of direct sticking and  reaccretion of rebounded or ejected aggregates due to gravity. \citet{Teiser2009b} showed that the size distribution of the ejected dust agglomerates does not vary significantly from the initial size distribution though fragmentation was observed upon
impact. A large amount of smaller fragments (few $\mu$m) form, but they contribute only 5 \% to the total fragment mass and therefore are of minor importance. As we study slower collisions here, we assume that this still holds. The velocity distributions for impact ejecta measured for the different impact velocities are shown in Fig. \ref{ejecta}.

The impact fragments are ejected from the impact site in arbitrary directions. As the ejecta velocities are derived from the images of only one camera only one component of the horizontal velocity of ejecta is measured. The complete horizontal velocity of the impact ejecta consists of two horizontal components and is therefore expected to be larger than the measured values by a factor of $\sqrt 2$, assuming that both horizontal components are the same. 

\begin{figure}[tb]
\includegraphics[width=7.5cm]{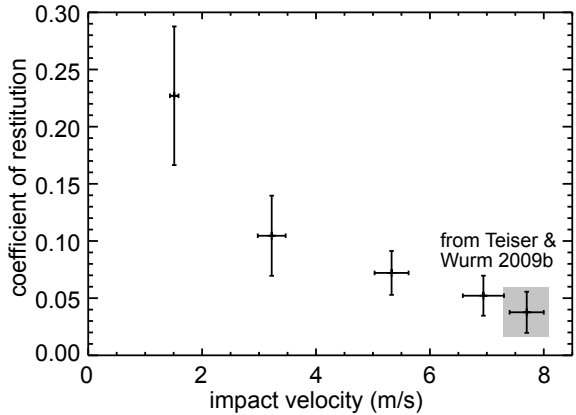}
\caption{Coefficient of restitution as determined from the mean ejecta  and impact velocities. The data point marked grey has been determined by \citet{Teiser2009b}}
\label{restitution}
\end{figure}

As most mass of the ejecta is in particles comparable in size to the impacting aggregates the collisions might be characterized by a coefficient of restitution, which we define as $\xi = v_{\mathrm{ejecta}}/v_{\mathrm{impact}}$. For this both horizontal components of the ejecta velociy are assumed to be the same on average due to symmetry. 

The velocities of the ejecta hardly change with increasing impact velocities, as already shown in Fig. \ref{ejecta}. This means that the coefficient of restitution decreases with impact velocity. The coefficient of restitution and its dependency on the collision speed is shown in \mbox{Fig. \ref{restitution}.} Here, the mean impact and ejecta velocities are used to calculate the coefficient of restitution and the errorbars are determined using the standard deviations for the velocities. The grey marked data point at \mbox{$v = 7.7$ m/s} is the result of the experiments by \citet{Teiser2009b}. With increasing impact velocities obviously more kinetic energy is dissipated, so the velocities of ejecta stay small. There are two processes which might be responsible for the increasing energy loss. One might be the formation of a large number of very small fragments as found in the size distribution by \citet{Teiser2009b}. However, these small fragments play only a minor role for the total mass of the particle. The other effect is internal restructuring of the impacted target, which should lead to a lower porosity of
the evolving aggregates with increasing collision velocity. 

\subsection{Porosities}\label{porosities}

\begin{figure}[tb]
\includegraphics[width=7.5cm]{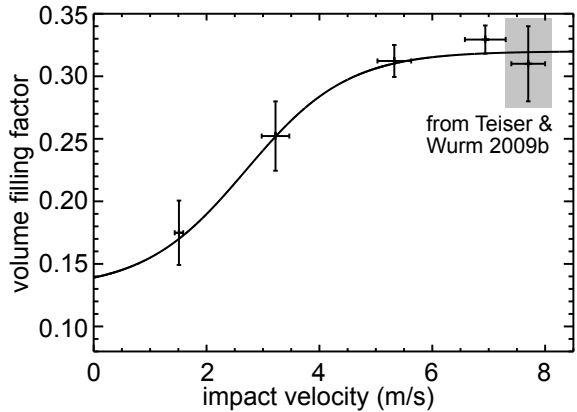}
\caption{Volume filling factor of the forming dust agglomerates and its dependency on the impact velocities. The data point marked grey has been determined by \citet{Teiser2009b}}
\label{porosity}
\end{figure}

The porosities have been determined for targets of (eight) different sizes, using images from different perspectives for the volume determination as already described in section \ref{methods}. The mean porosities or volume filling factor (1 - porosity) of the analyzed targets and their standard deviations for each impact velocity are shown in \mbox{Fig. \ref{porosity}}. The grey marked data point is the value determined by \citet{Teiser2009b}.

The volume filling factor of the forming dust aggregates increases with the impact velocity. This is expected, as the dynamic pressure of the colliding projectiles increases with the impact velocity. However, the volume filling factor saturates at impact velocities above 5 m/s and reaches a constant value of about $\phi = 0.32$. The plotted curve in \mbox{Fig. \ref{porosity}} is a function similar to the Fermi-function, which has already been used to charcterize the compression behaviour of dust agglomerates by \citet{Blum2006} and \citet{Guttler2009}. This function of the impact velocity $v_{imp}$ is given by
\begin{equation}
\phi(v_{imp}) = \phi_{max} - \frac{\phi_{max} - \phi_{min}} {\exp\left(\frac{v_{imp} - v_{mean}} {\Delta}\right) + 1}\,.
\end{equation}
Here, $\phi_{min}$ is the minimum porosity for dust samples prepared by sieving without any further compression and is set to $\phi_{min} = 0.13$, which is in agreement with previous studies by \citet{Paraskov2007}. The upper compression limit is set to $\phi_{max} = 0.32$ according to the experiments described here. This is close to the maximum value achieved by local compression as already proposed by \citet{Teiser2009a}. The other two parameters are adjusted to fit the curve to the data points. Here, $v_{mean}$ describes the velocity for which the mean value between $\phi_{min}$ and $\phi_{max}$ is reached and is found to be $v_{mean} = 2.7\, \mathrm{m/s}$. The other parameter describes the steepness of the transition between the minimum and the maximum volume filling and is found to be $\Delta = 0.9\, \mathrm{m/s}$. The compression with increasing impact velocity obviously is well represented by the used function.

\section{Discussion}\label{discussion}

The saturation volume filling factor is much larger than the volume filling factor feasible by uni-directional compression as determined by \citet{Blum2006}. In that case the maximum compression reachable with the dust material used in this work is $\phi = 0.20 \pm 0.1$. This can be understood as in our case the dust bed is much larger than the impacting aggregates and the dust bed surrounding the impact side provides support for higher compression at the impact side. However, at the same time higher filling factors are prohibited as the target gets destroyed in the close vicinity of the impact side if more pressure is applied which would lead to a pressure relief and keep the porosity at a constant level.

It is quite clear that higher compression (higher volume filling factors) in principle are possible, as e.g.  \citet{Guttler2009} measured a maximum volume filling of $\phi = 0.58$ but only if pressure is applied to all sides of a target. The impact experiments presented here suggest that the dynamic pressure induced by incoming small projectiles is neither uni-directional, nor omni-directional, but somewhere between. This is in agreement with the dust agglomerates used as targets and projectiles in the collision studies by \citet{Wurm2005b} and \citet{Teiser2009a}. 

The porosity in the exeriments by \citet{Kothe2010} produced volume filling factors up to $\phi = 0.4 \pm 0.06$ at 6 m/s. A difference and possibly a higher value for the saturation volume filling factor might result from the difference in monomer choice.  Spherical and monodisperse particles as used by \citet{Kothe2010} can be compressed more easily. This has e.g. been shown by experiments on uni-directional compression by \citet{Blum2006}. The limit in volume filling factor found by \citet{Weidling2009} for spherical particles was  $\phi = 0.36$. Within the error bars the largest value by \citet{Kothe2010} is still well above our saturation value. The porosity dependence on collision velocity in the data by \citet{Kothe2010} is steeper and they extrapolated the data to volume filling factors of $\phi = 0.6$ at about 10 m/s. Our experiments with irregular shaped grains, which would be expected in protoplanetary disks show, that the volume filling factor levels off at much lower values and lower velocities. We therefore regard such highly compressed targets as unlikely if an aggregate grows by addition of small aggregates and find that a value of $\phi = 0.32$ describes the volume filling factor of realistic arbitrarily shaped dust aggregates in a wide range. 

In the reported experiments the targets grow by a combination of direct sticking and reaccretion due to gravity. 
However, gravity in the laboratory replaces the gas drag in protoplanetary disks. Large dust aggregates move through the disk and feel a constant head wind corresponding to their drift and sedimentation speeds. Impact ejecta are exposed to this head wind and are accelerated back towards the target surface. This acceleration can be treated as constant to the first order and is given by $a = v/\tau$ with the wind speed $v$ and the coupling time $\tau$. The wind speed is equivalent to the collision velocitity. According to \citet{Blum1996}, the coupling time is given by 

\begin{equation}
\tau = 0.68 \frac{m}{A} \frac{1}{\rho_g\cdot \nu_g}\,. 
\end{equation}

Here, $\rho_g$ is the gas density, $\nu_g$ is the thermal velocity of the gas molecules 

given as

\begin{equation}
\nu_g = \sqrt{\frac{8 R_{\mathrm{gas}} T}{\pi \mu}}\,.
\end{equation}

Here, $R_{\mathrm{gas}}$ is the gas constant, $\mu = 2.34$ g/mol is the molar mass of the gas (mixture of hydrogen and helium), $T$ is the gas temperature, and $m/A$ is the ratio between particle mass and the cross section of the particle. For a spherical particle with a density $\rho_p$ and a volume filling factor $\phi$ it is given as $m/a = 4/3\cdot r \rho_p \phi$. To estimate the gas drag on the impact ejecta we use a volume filling of $\phi = 0.32$. The quartz dust used has a density of $\rho = 2.6$ g/cm$^3$. To estimate the acceleration by the head wind we consider the minimum mass solar nebula (MMSN) model by \citet{Hayashi1985}. The gas density is 

\begin{equation}
\rho_g = 1.4 \times 10^{-6} \left(\frac{R}{1\mathrm{AU}}\right)^{-11/4} \mathrm{kg/m^3}\,.
\end{equation}

and the temperature is

\begin{equation}
T = 280\left( \frac{R}{1 \mathrm{AU}}\right)^{-1/2} \mathrm{K}\,.
\end{equation}

For a particle of 100 $\mu$m radius with the given material parameters at 1 AU radial distance these equations lead to a coupling time of about \mbox{$\tau = 30$ s}. For the lowest collision velocity (head wind velocity) of 1.5 m/s this leads to an acceleration of \mbox{$a = 0.05$ m/s$^2$}, which is smaller than gravity in the laboratory by  a factor of 200. Therefore, the influence of reaccretion in a disk is reduced. Nevertheless, reaccretion even of large 100 $\mu$m fragements for cm-size aggregates in protoplanetary disk is still possible. This can be seen from simple estimates of the distance traveled by an ejected/rebounding aggregate before it hits a surface again with the a given acceleration. A particle launched from a surface with a vertical velocity of 0.1 m/s (see Fig. \ref{ejecta}) reaches the surface again after a time of $2 \cdot v/a$ or 4 s. If the particle travels with the same horizontal velocity it hits the surface a second time at a distance of 40 cm from the launch site.

In a MMSN collision velocities of 1.5 m/s correspond to a particle size of about 2 cm, according to \citet{Weidenschilling1993} and \citet{Sekiya2003}. Therefore, the considered particle will not be reaccreted but misses the target and is transported away. 
This might not support the idea of reaccretion but we started with the typical values on purpose to show that  even for the average particle the average acceleration is {\it only} a factor of 200 to low. We clearly note that the influence of reaccretion depends on the gas density, ejecta size and ejecta velocity.  The values above have been determined for the MMSN by \citet{Hayashi1985}. Other disk models, like e.g. developed by \citet{Brauer2008} or by \citet{Desch2007}, give slightly different values, as they use different density and temperature profiles for their disk models. For particles at 1 AU radial distance to the protostar the gas densities and temperature profiles given by \citet{Brauer2008} or by \citet{Desch2007} will lead to different coupling times. In case of the model by \citet{Brauer2008} no reaccretion by gas drag will occur, as the gas density in this model is very small and leads to a coupling time of $\tau = 1$ h. On the other hand, the disk model by \citet{Desch} gives a much denser disk and leads to a coupling time of $\tau = 1.2$ s for a dust agglomerate of $100 \mu$m in size at 1 AU. This is already sufficient for reaccretion, as the traveling distance is already reduced to 1.6 cm, which will already enable small bodies to reaccret material by gas drag. 

Additionally, the radial distance to the central star is also very important for the reaccretion process. In case of a radial distance of 0.5 AU, the coupling time of a 100 $\mu$m dust agglomerate is only 4.2 s in the MMSN by \citet{Hayashi1985}. This already reduces the traveling time of ejected fragments to 0.36 s, which leads to a travelled distance of 5.6 cm for a fragment of 100 $\mu$m in size. In case of smaller fragments (10 $\mu$m) this distance can easily be reduced to values of 5 mm. In this case reaccretion of impact ejecta will work more efficiently.

Growth has been observed on targets of different sizes with diameters in the range between 3.2 mm and 35 mm and the resulting volume filling is qualitatively the same. With a typical horizontal velocity of 0.1 m/s and a typical size of 100 $\mu$m, ejecta will travel 2 mm before lading on the target surface again. The horizontal velocities typically are larger than the measured horizontal comonent due to projection effects (only one camera). The assumption of 0.1 m/s is therefore rather low and ejecta will travel further than 2 mm. This leads to the conclusion that the amount of reaccreted particles is significantly smaller for the small (3.2 mm) targets than for the larger ones. Similar experiments by \citet{Teiser2009b} show similar results. Their experiments showed that aggregates growing on a thin glass edge (1.1 mm thick) have the same volume filling as agglomerates growing within a beaker or on the outer shell of a rotating cylinder. \citet{Teiser2009b} showed that there is no qualitative difference between complete reaccretion (beaker), small amount of reaccretion (glass edge) or no reaccretion (rotating cylinder).  It was shown in previous experiments \citep{Wurm2001a, Wurm2001b} that reaccretion of small particles does not only work in the molecular flow regime in which the streamlines are directed to the target surface, but also in the transition regime between the molecular flow regime and the hydrodynamic regime. Reaccretion was found to be an efficient  mechanism for growth of dust agglomerates.

The target shapes give evidence that direct sticking and reaccretion occur for all target sizes, as all grown targets have the same cone shape, independent of the target size or the impact velocity (Fig. \ref{targets}). \citet{Teiser2009b} developed a toy model for target growth by reaccretion. They showed that the probability for reaccretion is largest at the center of a target and becomes smaller to the edges. The result is a shape which is found in all our collision experiments of this type, as already shown in \mbox{Fig. \ref{targets}.} The distance travelled by ejecta before hitting a second time can be reduced easily by a factor of 20 by considering  either slightly smaller particles (projectile fragmentation), slightly smaller ejecta velocities, or slightly larger gas densities. The first two are present in the size and velocity distributions (\citet{Teiser2009b}, Fig. \ref{ejecta}) . The gas density is certainly higher and high enough for accreting almost all ejecta slighly inwards of 1 AU. So growth of cm aggregates to larger aggregates will occur if smaller aggregates are present.

\begin{figure}[tb]
\includegraphics[width=7.5cm]{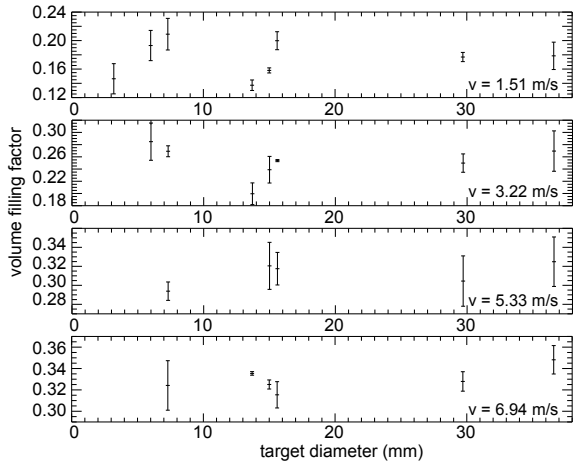}
\caption{Mean volume filling factor and its dependency on the target sizes for the four collision velocities.}
\label{thick}
\end{figure}

The influence of the target size and the ratio between directly sticking aggregates and reaccreted ejecta at different velocities on the resulting volume filling is analyzed in \mbox{Fig. \ref{thick}.} No significant dependency between the target size and the measured volume filling factor can be determined for any collision velocity, which is in agreement with the results by \citet{Teiser2009b}. The observed porosities therefore can be regarded as universal being independent of the target size but only depending on the collision velocity. 

\section{Conclusions}

Within this work the self consistent growth of macroscopic dust agglomerates by accretion of small dust aggregates has been analyzed. Experiments were carried out to bridge the size range between sub-mm particles and the decimeter size range, which is critical for a variety of growth models. In general we conclude that the reacccretion by gas drag and therfore the growth efficiency might vary strongly for different locations in disks, but that the growth of larger aggregates is possible in a growth model where sub-mm particles impact a growing aggregate. Together with earlier experiments by \citet{Teiser2009a} and \citet{Wurm2005b} we find that growth is possible up to collision velocities of at least 60 m/s.  

In detail, the specific experimental results of this work are:

\begin{itemize}
\item Dust agglomerates of a few centimeters in size or larger grow by a combination of direct sticking and reaccretion, if they are exposed to multiple impacts of sub-mm particles.
\item Reaccretion by gas drag can aid protoplanetary growth efficiently.
\item Large aggregates have a maximum volume filling of about $\phi = 0.32$. Although the evolving dust agglomerates become more compact when the impact velocities increase, a saturation is reached for impact velocities of more than 5 m/s.
\item The coefficient of restitution decreases with increasing impact velocities in the studied velocity regime up to about 8 m/s as the ejecta velocity
is essentially independent of the impact velocity. The energy is dissipated by compaction of the target surface and by projectile fragmentation.
 
\end{itemize}

\acknowledgments

We appreciate funding by the Deutsche Forschungsgemeinschaft within the frame of the research group FOR 759.


\clearpage


\begin{thebibliography}{}
\bibitem[Beitz et al. (2011)]{Beitz2011} Beitz, E., G\"uttler, C., Blum, J., Meisner, T., Teiser, J., Wurm, G. \ 2011, \apj, submitted.
\bibitem[Blum et al.(1996)]{Blum1996} Blum, J., Wurm, G., Kempf, 
S., \& Henning, T.\ 1996, \icarus, 124, 441
\bibitem[Blum 
\& Wurm(2000)]{Blum2000} Blum, J., \& Wurm, G.\ 2000, \icarus, 143, 138
\bibitem[Blum et al.(2006)]{Blum2006} Blum, J., Schr{\"a}pler, R., Davidsson, B.~J.~R., \& Trigo-Rodr{\'{\i}}guez, J.~M.\ 2006, \apj, 652, 1768
\bibitem[Blum 
\& Wurm(2008)]{Blum2008} Blum, J., \& Wurm, G.\ 2008, \araa, 46, 21
\bibitem[Brauer et 
al.(2008)]{Brauer2008} Brauer, F., Dullemond, C.~P., \& Henning, T.\ 2008, \aap, 480, 859 
\bibitem[Desch(2007)]{Desch2007} Desch, S.~J.\ 2007, \apj, 671, 
878
\bibitem[Dominik 
\& Tielens(1997)]{Dominik1997} Dominik, C., \& Tielens, A.~G.~G.~M.\ 1997, \apj, 480, 647
\bibitem[Goldreich \& Ward, 1972]{Goldreich1973} Goldreich, P. \& Ward, W. R. ,1973. \apj., 183, 1051–1062.
\bibitem[G{\"u}ttler et al.(2009)]{Guttler2009} G{\"u}ttler, C., 
Krause, M., Geretshauser, R.~J., Speith, R., 
\& Blum, J.\ 2009, \apj, 701, 130
\bibitem[G{\"u}ttler et 
al.(2010)]{Guttler2010} G{\"u}ttler, C., Blum, J., Zsom, A., Ormel, C.~W., \& Dullemond, C.~P.\ 2010, \aap, 513, A56
\bibitem[Hayashi et al.(1985)]{Hayashi1985} Hayashi, C., Nakazawa, 
K., \& Nakagawa, Y.\ 1985, Protostars and Planets II, 1100
\bibitem[Johansen 
\& Youdin(2007)]{Johansen2007a} Johansen, A., \& Youdin, A.\ 2007, \apj, 662, 627
\bibitem[Johansen et al.(2007)]{Johansen2007} Johansen, A., Oishi, 
J.~S., Mac Low, M.-M., Klahr, H., Henning, T., 
\& Youdin, A.\ 2007, \nat, 448, 1022
\bibitem[Kothe et al.(2010)]{Kothe2010} Kothe, S., G{\"u}ttler, 
C., \& Blum, J.\ 2010, \apj, 725, 1242
\bibitem[Paraskov et al.(2007)]{Paraskov2007} Paraskov, G.~B., Wurm, 
G., \& Krauss, O.\ 2007, \icarus, 191, 779
\bibitem[Safronov, 1969]{Safronov1969} Safronov, V. S. ,1969. NASA TTF.
\bibitem[Schr{\"a}pler \& Henning, 2004]{Schrapler2004} Schr{\"a}pler, R., \& Henning, T.\ 2004, \apj, 614, 960
\bibitem[Sekiya 
\& Takeda(2003)]{Sekiya2003} Sekiya, M., \& Takeda, H.\ 2003, Earth, Planets, and Space, 55, 263
\bibitem[Suyama et al.(2008)]{Suyama2008} Suyama, T., Wada, K., 
\& Tanaka, H.\ 2008, \apj, 684, 1310 
\bibitem[Teiser 
\& Wurm(2009a)]{Teiser2009a} Teiser, J., \& Wurm, G.\ 2009, \mnras, 393, 1584
\bibitem[Teiser 
\& Wurm(2009b)]{Teiser2009b} Teiser, J., \& Wurm, G.\ 2009, \aap, 505, 351 
\bibitem[Wada et al.(2008)]{Wada2008} Wada, K., Tanaka, H., 
Suyama, T., Kimura, H., \& Yamamoto, T.\ 2008, \apj, 677, 1296
\bibitem[Weidenschilling(1997)]{Weidenschilling1997} Weidenschilling, 
S.~J.\ 1997, \icarus, 127, 290
\bibitem[Weidenschilling 
\& Cuzzi(1993)]{Weidenschilling1993} Weidenschilling, S.~J., \& Cuzzi, J.~N.\ 1993, Protostars and Planets III, 1031
\bibitem[Weidling et al.(2009)]{Weidling2009} Weidling, R., 
G{\"u}ttler, C., Blum, J., \& Brauer, F.\ 2009, \apj, 696, 2036
\bibitem[Wurm et al.(2005a)]{Wurm2005a} Wurm, G., Paraskov, G., 
\& Krauss, O.\ 2005, \pre, 71, 021304
\bibitem[Wurm et al.(2005b)]{Wurm2005b} Wurm, G., Paraskov, G., 
\& Krauss, O.\ 2005, \icarus, 178, 253
\bibitem[Wurm et al.(2001a)]{Wurm2001a} Wurm, G., Blum, J., 
\& Colwell, J.~E.\ 2001, \icarus, 151, 318 
\bibitem[Wurm et al.(2001b)]{Wurm2001b} Wurm, G., Blum, J., 
\& Colwell, J.~E.\ 2001, \pre, 64, 046301 
\bibitem[Youdin 
\& Johansen(2007)]{Youdin2007} Youdin, A., \& Johansen, A.\ 2007, \apj, 662, 613 
\bibitem[Zsom et 
al.(2010)]{Zsom2010} Zsom, A., Ormel, C.~W., G{\"u}ttler, C., Blum, J., \& Dullemond, C.~P.\ 2010, \aap, 513, A57
\end{thebibliography}
\end{document}